\documentclass[graybox]{svmult}

\usepackage[numbers,sort]{natbib}

\usepackage{mathptmx}       
\usepackage{helvet}         
\usepackage{courier}        
\usepackage{type1cm}        
\usepackage{makeidx}         
\usepackage{graphicx}        
\usepackage{multicol}        
\usepackage[bottom]{footmisc}

\def\apj{ApJ }
\def\aj{AJ }

\def\lb{L_{\rm B}}
\def\lx{L_{\rm X}}

\def\vr{v_{\rm rot}}
\def\sgc{\sigma _{\rm c}}

\catcode`\@=11
\def\gsim{\ifmmode{\mathrel{\mathpalette\@versim>}}
    \else{$\mathrel{\mathpalette\@versim>}$}\fi}
\def\lsim{\ifmmode{\mathrel{\mathpalette\@versim<}}
    \else{$\mathrel{\mathpalette\@versim<}$}\fi}
\def\@versim#1#2{\lower 2.9truept \vbox{\baselineskip 0pt \lineskip
    0.5truept \ialign{$\m@th#1\hfil##\hfil$\crcr#2\crcr\sim\crcr}}}
\catcode`\@=12

\makeindex             

\begin{document}

\title*{Hot gas flows on global and nuclear galactic scales}
\titlerunning{Hot gas flows}
\author{Silvia Pellegrini}
\institute{Silvia Pellegrini \at Department of Astronomy, University of 
Bologna, via Ranzani 1, 40127 Bologna, Italy, \email{silvia.pellegrini@unibo.it}}
%
%
\maketitle

\abstract{ Since its discovery as an X-ray source with the $Einstein$
  Observatory, the hot X-ray emitting interstellar medium of
  early-type galaxies has been studied intensively, taking advantage
  of observations of improving quality performed by the subsequent
  X-ray satellites $ROSAT$, $ASCA$, $Chandra$ and $XMM-Newton$, and
  comparing the observational results with extensive modeling by means
  of numerical simulations.  The hot medium originates from the
  ejecta produced by the normal stellar evolution, and during
  the galaxy lifetime it can be accumulated or expelled from the
  galaxy potential well.  The main features of the hot gas evolution
  are outlined here, focussing on the mass and energy input rates, the
  relationship between the hot gas flow and the main properties
  characterizing its host galaxy, the flow behavior on the nuclear and
  global galactic scales, and the sensitivity of the flow to major
  galaxy properties as the shape of the mass distribution and the mean 
  rotation velocity of the stars.}

\section{Introduction} \label{intro} 

X-ray observations performed during the late 1970's first revealed
that early-type galaxies (ETGs) emit soft thermal X-rays and then host
an interstellar medium (ISM), of a mass up to $\sim
10^{10}\,M_{\odot}$, that had not been discovered previously at other
frequencies because of its high temperature ($T\sim 10^7$ K; see
Fabbiano, this volume).  Such a medium was the long sought phase fed
by stellar mass losses and predicted by the stellar evolution studies
(e.g., \citep{FG,Mat89}); in fact, both isolated ETGs and those in
groups or clusters emit soft thermal X-rays \cite{Fab89}, which
provides simple evidence that the hot ISM is mostly indigenous rather
than accreted intergalactic medium.  Besides solving a puzzle, this
discovery opened the study of an important component of ETGs: it is a
major ingredient of galactic evolution, see for example its role in
feeding a central supermassive black hole and maintaining an
activity cycle and starformation (e.g., \citep{FC88,For05,Kor}, and Ciotti \& Ostriker, this
volume), or that in polluting the space surrounding
ETGs with metals, via galactic outflows (\cite{R93}, and Pipino, this
volume), or finally in responding to environmental
effects, as interaction with neighbors, stripping, sloshing, and
conduction (e.g., \citep{Acr,Sun07,K08}, and Sarazin, this volume).

The most striking X-ray observational feature of ETGs is the wide
variation in their luminosity ($\lx$) values, of $\gsim 2$ orders of
magnitude at any fixed galactic optical luminosity $L_B$, when
$L_B>3\times 10^{10}L_{B,\odot}$ \citep[][]{Fab89,OS,Mem, Kim10}.
This feature cannot be explained by distance uncertainties, since a
variation of the same size is present even in the distance-independent
diagram of $\lx/\lb$ versus the central stellar velocity dispersion
$\sigma_c$ \cite[e.g.,][]{Esk}.  A number of reasons have been
proposed as responsible for this large variation, of environmental
nature (see Sarazin, this volume) or linked to the possibility for the
gas content to evolve substantially during the ETG lifetime. In this
latter context, many studies investigated with numerical simulations
the dynamical evolution of the hot ISM in ETGs
\citep[][]{SW,LM,Da90,Da91,C91,PC98,T09}.  In more recent times,
the effect of feedback from a central supermassive black hole (MBH)
has revealed as another potential contributor to the variation of the
hot ISM luminosity (see Statler, and Ciotti \& Ostriker, this volume).

Below I briefly review our current knowledge about the feeding and the
energetics of the hot gas flows, concentrating on their dynamical
state as a function of galactic mass and other major galaxy
properties.  Section~\ref{gasfl} gives an updated summary of the
fundamental elements entering the problem, the mass and energy inputs to the
flow; Sect.~\ref{dec} presents the general case for the evolution of
the flow, pointing out the different behavior on the global and
nuclear galactic scales, and making also use of a representative ETG;
the effect of a central MBH, acting as a gravitating point mass, and
the expectations for accretion feedback, are presented in
Sect.~\ref{mbh}; finally, Sect.~\ref{sensit} discusses the observed
sensitivity of the flow to major galaxy properties as the flattening of the
mass distribution, the mean rotational velocity of the stars, and the shape of the
stellar profile.

\section{Feeding and Energetics of the Hot Gas Flows}\label{gasfl}

In this Section the fundamental processes and quantities at the basis
of the origin and evolution of hot gas flows are introduced: their
feeding via stellar mass losses (Sect.~\ref{maslos}), their heating
via type Ia supernovae explosions (Sect.~\ref{sna}), and their
energy budget (Sect.~\ref{enbud}).

\subsection{The Stellar Mass Loss Rate}\label{maslos}

In ETGs the gas is lost by evolved stars mainly during the red giant,
asymptotic giant branch, and planetary nebula phases.  These
losses originate ejecta that initially have the velocity of the parent
star, then interact with the mass lost from other stars or
with the hot ISM, and mix with it.  The details of the interaction are
controlled by several parameters, like the velocity of the mass loss
relative to the hot phase, or the density of the ambient ISM
\cite{Mat90}.  Parriott \& Bregman (\cite{PB} and \cite{BP}), modeling
the interaction with two-dimensional hydrodynamical simulations, found
that most of the continuous mass loss from giant stars is heated to
approximately the temperature of the hot ISM within few parsecs of the
star; in the case of mass ejected by planetary nebulae, about half of
the ejecta separates and becomes hot, and the other half creates a
narrow wake that remains mostly cool, unless turbulent mixing allows
for its heating on larger scales.  Far infrared observations allow us
to measure directly the stellar mass loss rate for the whole galaxy
($\dot M_*$); this was
for example derived for nine local ETGs from $ISO$ data
\cite{At}. When rescaled by the luminosities of the respective
galaxies, the values of $\dot M_*$ were found to vary by a factor of
$\sim 10$, which was attributed to different ages and metallicities.
The average of the observed rates was $\dot M_*=7.8\times 10^{-12}\,\,
L_B(L_{B,\odot})$ M$_{\odot}$yr$^{-1}$, and was found to be in
reasonable agreement with previous theoretical predictions \cite{At}.

\begin{figure}[b]
\vskip -2truecm
\includegraphics[scale=.4]{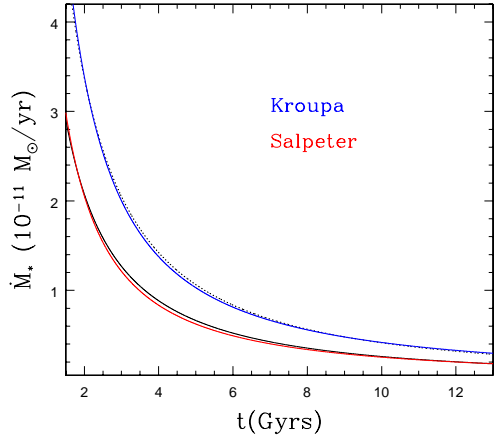}
\vskip -0.5truecm
\caption{Evolution of the stellar mass loss rate $\dot M_*(t)$
according to the models in \cite{Mar05}, for a Salpeter IMF (solid
line with best fit of Eq.~\ref{eq:mdots} in red) and for a Kroupa IMF
(dotted line with best fit of Eq.~\ref{eq:mdots} in blue). Solar
abundance is assumed, and a total stellar mass of 1 $M_{\odot}$ at an
age of 12 Gyrs.}  
\label{fig1} 
\end{figure}

According to single burst stellar population synthesis models, most of
the stellar mass is lost at early times, before an age of $\sim 2$
Gyr. For example, the mass lost by stars at an age of 2 Gyr is 25\% of
the total initial mass, for the Salpeter initial mass function (IMF), and
36\% for the Kroupa IMF; at an age of 12 Gyr, the additional loss is
of $\sim 4$\% and 6\% of the initial mass, for the same
two IMFs respectively \cite{Mar05}.  Figure~\ref{fig1} shows the trend of $\dot
M_*(t)$ with time, estimated from the models in \cite{Mar05} for
solar metal abundance. At an age of $\gsim 2$ Gyr, this trend can be
approximated as:
\begin{equation}
\dot M_* (t) = 10^{-12}\, A\,\times M_*(M_{\odot})\,\, t_{12}^{-1.3} 
\quad\quad\quad ({\rm M_{\odot} yr^{-1}}),
\label{eq:mdots}
\end{equation}
where $M_*$ is the galactic stellar mass\footnote{The stellar mass $M_*$ 
changes very little at late epochs, for example by $<1$\% for a variation of 
$\pm 2$ Gyr at an age of 12 Gyr.} at an age of 12 Gyr, 
$t_{12}$ is the age in units of 12 
Gyrs, and $A=2.0$ or 3.3 for a Salpeter or Kroupa IMF (see Fig.~\ref{fig1}).

The relation above agrees well with previous theoretical estimates
\cite{Mat89,C91}.  Taking the stellar mass-to-light ratio in the
B-band at an age of 12 Gyr ($M_*/L_B=9.17$ and 5.81, respectively for
the Salpeter and Kroupa IMFs, \cite{Mar05}), Eq.~\ref{eq:mdots}
gives $\dot M_* $(12 Gyr)= B $\times 10^{-11} \,
L_B(L_{B,\odot})$ M$_{\odot}$yr$^{-1}$, with B=1.8 or B=1.9 for the
Salpeter or Kroupa IMF.  The latter relation gives a rate that is
roughly double as large as the average of the observational estimates
quoted above (\cite{At}), that however has a large variation
around it, partly explained by differences in the ages and
metallicities of the observed ETGs.

\subsection{The Type Ia Supernovae Mass and Energy Input}\label{sna}

The total mass loss rate of a stellar population $\dot M$ is given by
the sum $\dot M(t)=\dot M_*(t)+ \dot M_{\rm SN}(t)$, where one adds to
$\dot M_*(t)$ (discussed in the previous Sect.~\ref{maslos}) 
the rate $ \dot M_{\rm SN}(t)$ of mass lost by type Ia supernovae (SNIa) events, 
the only ones observed in an old stellar population \cite[e.g.,][]{Capp}.  
The mass input due to SNIa's is $\dot M_{\rm
SN}(t)=1.4$ M$_{\odot}\, R_{\rm SN}(t)$ M$_{\odot}$ yr$^{-1}$, where $R_{\rm
SN}(t)$ (in yr$^{-1}$) describes the evolution of the explosion rate with 
time, since each SNIa ejects 1.4 M$_{\odot}$. In a detailed scenario for the 
SNIa precursors and their subsequent explosion past a burst of star
formation \citep{Gre05,Gre10}, $R_{\rm SN}(t)$ experiences a raising
epoch during the first 0.5-1 Gyr, at the end of which it reaches a peak, and
then decreases slowly with a timescale of the order of 10 Gyr, and
accounts for the present day observed rate.  A parameterization
of the rate after the peak, in number of events per year, is
\begin{equation}
R_{\rm SN}(t)=0.16  (H_0/70)^2 \times 10^{-12} \,\,
L_B (L_{B,\odot}) \,\,\, t_{12}^{-s}\quad\quad
         ({\rm yr}^{-1}) ,
\label{eq:rsn}
\end{equation}
where $H_0$ is the Hubble constant in units of km s$^{-1}$ Mpc$^{-1}$,
$L_B$ is the present epoch galaxy luminosity, and $s$ describes the
past evolution; when $t_{12}=1$, Eq.~\ref{eq:rsn} gives the rate for
local ETGs in \cite{Capp}, that has an uncertainty of $\pm 30$\%. 
Recently, new measurements of the observed rates of supernovae in the
local Universe, determined from the Lick Observatory Supernova Search
(LOSS; \cite{Li}), gave a SNIa's rate in ETGs
consistent with that in \cite{Capp}.  For the rate in
Eq.~\ref{eq:rsn}, and for $H_0=70$ km s$^{-1}$ Mpc$^{-1}$, one obtains
$\dot M_{\rm SN}$(12 Gyr) =2.2$ \times 10^{-13} L_B(L_{B,\odot}) $
M$_{\odot}$ yr$^{-1}$, that is almost $\sim 100$ times smaller than
the "normal" stellar mass loss rate $\dot M_* $ (12 Gyr)$\approx
2\times 10^{-11} \, L_B(L_{B,\odot})$ M$_{\odot}$ yr$^{-1}$ derived above
(Sect.~\ref{maslos}).

\begin{figure}[b]
\vskip -3truecm
\hskip -0.5truecm
\includegraphics[scale=.5]{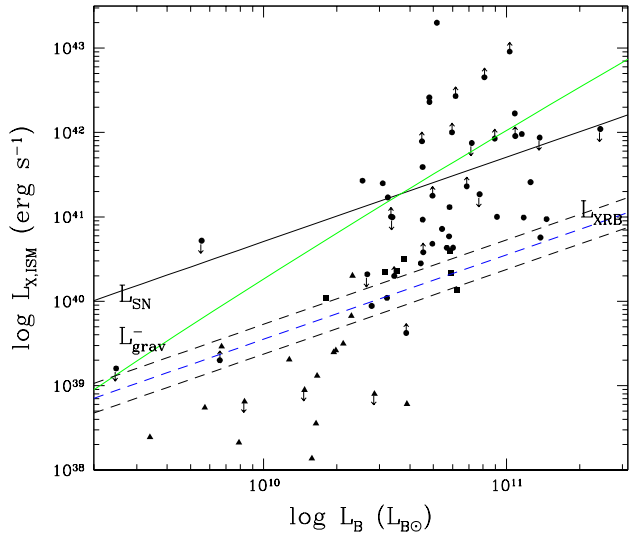}
\vskip -0.5truecm
\caption{The X-ray luminosity of the hot ISM $L_{X,ISM}$ 
from $Chandra$ observations, versus the
galactic $L_B$ (from corrected apparent magnitudes from
Hyperleda, and distances from \cite{T01,DS1}, for $H_0=70$
km s$^{-1}$ Mpc$^{-1}$). $L_{X,ISM}$
is from \cite{DS1} (circles), \cite{D06} (triangles), \cite{NM}
(squares); downward (upward) arrows indicate upper (lower) limits. The
dashed lines show the mean total luminosity from X-ray binaries
derived from $Chandra$ data (in blue), and its $\pm 1\sigma$
uncertainty, from \cite{Kim10}. All X-ray luminosities have been
recalculated for the 0.3--5 keV band, assuming the spectral shape in
the original references. The solid line is $L_{SN}$ 
(Eq.~\ref{eq:lsn}); the green line is $L_{grav}^-$ (Eq.~\ref{eq:lm}),
calculated for $\dot M_*(12$ Gyrs) of the Kroupa IMF
(Sect.~\ref{maslos}), and representative galactic mass models (see
also Sect.~\ref{dec}) with: a stellar de Vaucouleurs profile, $\sigma_c$
from the Faber-Jackson relation, an effective radius $R_e$ from the
Fundamental Plane relation (\cite{Ber}), a NFW dark mass profile with
a dark-to-luminous mass ratio of 5 in total  (\cite{kom}), and 
of 0.6 within $R_e$ (\citep{Cap06,Wei09,Shen}), a ratio of 
the scale radii of the dark and stellar mass distributions of
$r_{h}/R_e=1.5$.}
\label{f2} 
\end{figure}

SNIa's provide also heating (Sect.~\ref{enbud} below) at a rate
$L_{\rm SN}(t)$ that is the product of the kinetic
energy injected by one event ($E_{\rm SN}\approx 10^{51}$ erg) times
the rate $R_{\rm SN}(t)$.  This assumes that the total of
$E_{\rm SN}$ is turned into heat of the hot ISM, an assumption that is
clearly an overestimate, but not totally unreasonable for the hot
diluted gas (\cite{Mat89}).  Then
\begin{equation}
L_{\rm SN}(t) = E_{\rm SN}\times R_{\rm SN}(t) = 
5.1  (H_0/70)^2 \times 10^{30}\, 
L_B (L_{B,\odot})\,\, t_{12}^{-s}\quad\quad ({\rm erg \,\, s}^{-1}),
\label{eq:lsn}
\end{equation}
and is plotted in Fig.~\ref{f2} for $t_{12}=1$ and $H_0=70$ km
s$^{-1}$ Mpc$^{-1}$.  The SNIa's specific heating, given by the total
SNIa's heating per total injected mass (approximated hereafter with $\dot M_*$) is
\begin{equation}
{L_{\rm SN}\over {\dot M_*}}=1.6(H_0/70)^2 \times 10^{50}\, {L_B(L_{B,\odot})
\over A\, M_*(M_{\odot})} \, t_{12}^{1.3-s}
 \,\,\,\,\, ({\rm erg \,\, M_{\odot}^{-1}})
\label{eq:lsnspec}
\end{equation}
where $\dot M_*$ given in Eq.~\ref{eq:mdots} has been used. At an age
$t_{12}=1$, for $H_0=70$ km s$^{-1}$ Mpc$^{-1}$, and using $\dot M_*$(12 Gyr)
$\approx 2\times 10^{-11} \,L_B (L_{B,\odot})$ M$_{\odot}$ yr$^{-1}$ (valid 
for both IMFs, see below Eq.~\ref{eq:mdots}), one gets
$L_{\rm SN}/{\dot M_*}\approx 8\times 10^{48}\,\, {\rm erg \,\,
M_{\odot}^{-1}}$, that is a significant heating (with respect to, e.g., 
the specific binding energy of the gas, see Sect.~\ref{enbud} and 
Fig.~\ref{f2}).

During the galaxy lifetime, $L_{\rm SN}/{\dot M_*} \propto t^{1.3-s}$,
and then can increase or decrease with time increasing, with
consequences on the secular gas flow behavior (\citep{LM,Da90,C91};
Sect.~\ref{dec}).  Early hydrodynamical simulations of hot gas flows
enlighted the importance of the cosmological evolution of the SNIa's
rate to avoid excessive, and unobserved, mass accumulation at the
galactic centers: if $L_{SN}$ decreases faster than $\dot
M_*$ ($s>1.3$), then at early times it can be large enough to drive
the gas lost by stars in a supersonic wind, that later can become
subsonic and evolve into an inflow (\cite{C91}; Sect.~\ref{enbud}
below).  Subsequently, mainly following $Chandra$ and $XMM-Newton$
observations, it was realized that another major source of ISM heating
can be provided by the central MBH. The MBH heating is not sufficient
by itself to avoid long-lasting and massive inflows at early times,
but when coupled with the SNIa's heating, it can 
help the galaxy degassing and prevent large mass
accumulation at the galactic centers, independently of the relative
rates of $\dot M_*(t)$ and $R_{\rm SN}(t)$ (see Ciotti \& Ostriker,
this volume). Recent estimates of the slope $s$ agree with
a value around $s\sim 1$ (\citep{Man05,Gre05,Gre10,Maz10,Sha10}).

\subsection{Energetics of the Gas Flows}\label{enbud}

The material lost by stars is ejected at a velocity of few tens of km
s$^{-1}$ and at a temperature of $\lsim 10^4$ K (\cite{PB}); it is 
then heated to X-ray emitting temperatures by
the thermalization of the stellar velocity dispersion (as it collides
with mass lost from other stars or with the ambient hot gas, and is
shocked) and of the kinetic energy of SNIa's events (Sect.~\ref{sna}).  
The first process heats the ISM at a rate
\begin{equation}
L_{\sigma} = {1\over 2}{\dot M_*(t)\over M_*}
\int ^{\infty}_{0}4\pi r^{2}\rho_*(r) \sigma^{2}(r)dr,
\label{eq:lsig}
\end{equation}
where $\rho_*(r)$ is the stellar density profile, and 
$\sigma (r)$ is the trace of the local stellar velocity dispersion tensor.
The latter can be obtained by solving the
Jeans equations for an adopted galaxy mass model (e.g., \cite{bt87}), 
and for an assumed stellar orbital anisotropy. So doing, one derives that
the stellar heating $L_{\sigma}$ is a few times lower than that provided
by SNIa's, $L_{SN}$, for reasonable stellar and 
dark matter distributions (\cite{C91}; see also Sect.~\ref{test}).

In case of gas flowing to the galactic center (inflow), as the gas
falls into the potential well, power is generated that
can heat the gas; for a steady inflow of $\dot M_*$ through
the galactic potential down to the galactic center, this
power is given by
\begin{equation}
L_{grav}^+ ={\dot M_*(t)\over M_*}\int ^{\infty}_{0}4\pi r^{2}\rho_*(r)
[\phi (r) - \phi(0)] dr
\label{eq:lp}
\end{equation}
where $\phi (r)$ is the total potential [and the equation above applies 
 to mass distributions with a finite value of $\phi(0)$].
In outflows, the work  done against the gravitational field to extract steadily
and bring to infinity the gas shed per unit time $\dot M_*(t)$ is
\begin{equation}
L_{grav}^-=-{\dot M_*(t)\over M_*} \int ^{\infty}_{0}4\pi r^{2}\rho_*(r)
\phi(r) dr.
\label{eq:lm}
\end{equation}
$L_{\rm grav}^{-}$ is the minimum power required to steadily remove
the stellar mass loss, because of energy losses due to cooling, but these are
expected to be low in the low density of an outflow. Figure~\ref{f2}
shows an example of $L_{grav}^-$ as a function of $L_B$ for 
representative two-component galaxy mass models.

Even though stationary conditions are unlikely to be verified, the
quantities above are useful to evaluate in first approximation the
energy budget of the flow, and then predict its dynamical state and the
gas content of an ETG. In a more compact notation, the integral in
Eq.~\ref{eq:lm} can be expressed as
\begin{equation}
L_{grav}^-=\dot M_*(t)\sigma_c^2 \Gamma^- (\cal{R},\beta)
\label{eq:lmbis}
\end{equation}
where $\sigma_c$ is the central stellar velocity dispersion, and the
dimensionless function $\Gamma^-$ depends on the depth and shape of the
potential well, via the variables $\cal{R} $=$M_h/M_*$ ($M_h$ being the total
dark halo mass), and $\beta=r_{h}/r_{*}$ ($r_{h}$ and $r_{*}$ being the
scale radii of the dark and stellar mass distributions; see
\citep{C91,CP, PC98} for examples of $\Gamma^- $ for various stellar
density profiles). $\Gamma^-$ increases for increasing ${\cal R}$ and decreasing
$\beta$ (e.g., \cite{P11})\footnote{An expression equal to that in Eq.~\ref{eq:lmbis}
can be written for $L_{grav}^+$, just replacing $\Gamma^-$ with a function 
$\Gamma^+ ({\cal R},\beta)$; the latter  has the
same trend as $\Gamma^-$ to increase for increasing ${\cal R}$ and decreasing
$\beta$; for reasonable galaxy mass models,  $\Gamma^+ \sim 2\Gamma^-$
(\cite{P11}).}. 
For reasonable galaxy structures, $\Gamma^-$ does
not largely vary with $\cal R$ and $\beta$, as for example empirically
demonstrated by the existence of scaling laws as the Fundamental Plane
of ETGs (\cite{RC93}).  Equation~\ref{eq:lmbis} then shows
how the larger $\sigma_c $, the harder for the gas to leave the
galaxy; for example, per unit gas mass, one has $L_{grav}^-/\dot M_* 
\propto \sigma_c^2 $.
$L_{grav}^-$ is expected to steeply
increase with $L_B$, since on
average the larger is $\sigma_c $, the brighter is the optical luminosity
of the ETG (from the Faber-Jackson relation $L_B\propto \sigma_c^4$), 
and since also $\dot M_*$ is proportional to $L_B$
(Sect.~\ref{maslos}).  For a fixed galaxy structure [i.e., a fixed
$\Gamma^- (\cal{R},\beta$)], and using the Faber-Jackson relation, one has
$L_{grav}^- \propto L_B^{1.5}$, a trend close to 
the green line in Fig.~\ref{f2} (that has a slope of $\sim 1.8$, though, since it
derives from the full $L_B-\sigma_c$ relation for ETGs, that has a
shallower slope at the lower $L_B$, \cite{Dav}).  Since the available
heating to extract the gas ($L_{SN}$) increases with $L_B$ as well, it
is useful to evaluate the run of the ratio between the power required
to extract the gas ($L_{grav}^-$) and that given by SNIa's\footnote{In
order to account for all the heating sources, one should add to the
denominator of Eq.~\ref{eq:chi} also $L_{\sigma}$, that has been
neglected for simplicity, since typically $L_{\sigma}<<L_{SN}$, as
written below Eq.~\ref{eq:lsig}.}: 
\begin{equation} {L_{grav}^-
(t)\over L_{\rm SN}(t)}\propto t_{12}^{s-1.3}\sigma_c ^2
             \Gamma^- (\cal{R},\beta).
\label{eq:chi}
\end{equation}
In a first approximation, (most of) the galaxy will host an outflow if
this ratio has always been lower than unity, and an inflow soon after
it becomes larger than unity (\cite{C91}). The time evolution of the ratio is
determined by the value of $s-1.3$; recent progress indicates $s\gsim 1$
(end of Sect.~\ref{sna}), which produces a ratio in
Eq.~\ref{eq:chi} decreasing with time. This means that with time
increasing the gas has a tendency to become hotter and, if outflowing
regions are present, the degassing becomes faster (see also
Sect.~\ref{test}). Equation~\ref{eq:chi} also indicates the underlying
cause of the average $L_{X,ISM} -L_B$ correlation (Fig.~\ref{f2}), that is the increase of $L_{grav}^-/L_{\rm
SN}$ with $\sigma_c ^2$, and then with $L_B$
(provided that the function $\Gamma^-$ does not vary widely with $L_B$,
as expected for reasonable galaxy mass models).

The relative size of $L_{grav}^- $ and $L_{\rm SN}$ can be estimated
from Fig.~\ref{f2}. $L_{grav}^- < L_{SN} $ for $L_B\lsim 3\times
10^{10}L_{B,\odot}$, therefore in these ETGs we can expect outflows to
be important, and then low $L_{X}$ values. This "prediction" has been
confirmed to be true recently, thanks to $Chandra$ observations of low
$L_B$ galaxies, probing for the first time gas emission levels even
below those of the X-ray binaries emission
(\citep{D06,P07,T08,Kim10}).  
For $L_B>3\times 10^{10}L_{B,\odot}$, instead,
$L_{grav}^- > L_{\rm SN}$ and the SNIa's heating is insufficient to
prevent (at least some) inflow.  At high $L_B$, however, a very large
variation of $L_{X}$ is observed, from values typical of winds to
values even larger than predicted by models for global inflows in
isolated ETGs (see Sect.~\ref{dec} below). This has a few possible
explanations: on one hand, there is the high sensitivity of the gas
behavior to variations in the parameters entering Eq.~\ref{eq:chi}
(the dark and stellar mass, their distribution, the orbital structure,
and possibly also $L_{SN}$, can all vary at fixed $L_B$), as discussed
in Sects.~\ref{dec} and~\ref{sensit} below (see also
\citep{C91,PC98}). On the other hand, the simple arguments above do
not consider important factors that can influence the hot gas content,
as injection of energy from the nucleus (see Ciotti \& Ostriker, this
volume; \cite{PCO}), and/or environmental effects (Sarazin, this volume;
\citep{P99a,BBreg,BM98,Hel,Sun07,Jel,Sun09}). The considerations in
this Section can account for the average trend of $L_{X,ISM}$ with
$L_B$, but significant effects can be superimposed by these factors.

\section{Decoupled Flows and variation in $L_{X,ISM}$}\label{dec}

See the full chapter (chapter 2 in {\it Hot Interstellar Matter in
Elliptical Galaxies},  Springer, 2012;
http://www.springer.com/astronomy/book/978-1-4614-0579-5)

\subsection{Gas Temperature and Galaxy Structure}\label{tgas}

In the simulations described above (Sect.~\ref{dec}), the average
emission weighted gas temperature $T_{gas}$ ranges between 0.3 and 0.8
keV, for a large set of galaxy models with different $L_B$, dark
matter fraction and distribution, and SNIa's rate. This range of
$T_{gas}$ compares well with that of the gas temperatures recently
determined using $Chandra$ data (e.g., \cite{DS2,NM,Kim10}). Both in
the observational results and in the models $T_{gas}$ shows a trend to
increase with $\sigma_c$. For example, the statistical analysis of the gas temperature for
a sample of luminous ETGs observed with $ROSAT$ indicates
a correlation of type $\sigma_c\propto T_{gas}^{0.56\pm 0.09}$, although
with a large degree of scatter about this fit (\cite{OPC}).
Both the size of the $T_{gas}$ values and their trend with $\sigma_c$
behave as expected, since the gas temperature cannot be much
different from the virial temperature $T_{vir}$ of the galaxy
potential well.  In fact $T_{vir}$ is defined as
\begin{equation}
T_{vir}={1\over 3k} {\mu m_p \over M_*} \int 4 \pi r^2 \rho_*(r) \sigma^2 (r) \,dr 
\label{tvir}
\end{equation}
where $k$ is the Boltzmann constant, $\mu m_p$ is the mean particle
mass, $m_p$ is the proton mass, and $\sigma (r)$ is the
three-dimensional velocity dispersion (as in Eq.~\ref{eq:lsig}). As already
done for $L_{grav}^+$ and $L_{grav}^-$ in Sect.~\ref{enbud}, and in
analogy with Eq.~\ref{eq:lmbis}, $T_{vir}$ can be expressed
as $T_{vir}=\mu m_p\, \sigma_c^2 \Omega ({\cal R},\beta)/k$,
with $\Omega <1$. $T_{vir}$ is then
proportional to $\sigma_c^2$, which explains the trend of $T_{gas}$ with
$\sigma_c$ present in the models, and is close to the trend shown by
the observations (\cite{OPC}).
A simplified version of the virial temperature in
Eq.~\ref{tvir} is often used, i.e., $T_{\sigma}=\mu m_p\sigma_c^2/k$; this
somewhat overestimates the true $T_{vir}$, since $\Omega <1$.
One can notice that $T_{vir}$ is also the temperature linked to the
gas heating provided by the thermalization of the stellar random motions
(Eq.~\ref{eq:lsig}). Therefore  the values of $T_{vir}$ are
expected to represent a lower boundary to the values of $T_{gas}$,
due to the importance of additional heating mechanisms (as that 
provided by SNIa's). 

(abridged)

\subsection{Reasons for Decoupling}\label{reasons}

See the full chapter (chapter 2 in {\it Hot Interstellar Matter in
Elliptical Galaxies},  Springer, 2012;
http://www.springer.com/astronomy/book/978-1-4614-0579-5)

\subsection{The Gas Flow in a Testcase ETG }\label{test}

This Section presents the evolution and the properties of the flow for
a testcase ETG, whose optical properties place it where the
$L_{X,ISM}$ variation is of $\sim 2 $ orders of magnitude
(Fig.~\ref{f2}): $L_B=5\times 10^{10}L_{B,\odot}$ and $\sigma_c=260$
km s$^{-1}$, from the Faber-Jackson relation; the stellar mass profile
follows a S\'ersic law with index $n=5$, as
appropriate for the chosen $L_B$ (e.g., \cite{Kor}); the effective
radius $R_e=6.5$ kpc, from the Fundamental Plane relation (\cite{Ber}). The dark
halo has a NFW profile with $\cal{R}$=4 and a scale radius larger
than that of the stars ($\beta =r_h/R_e=1.5$). 

(abridged)

\begin{figure}[t]
\includegraphics[scale=.45]{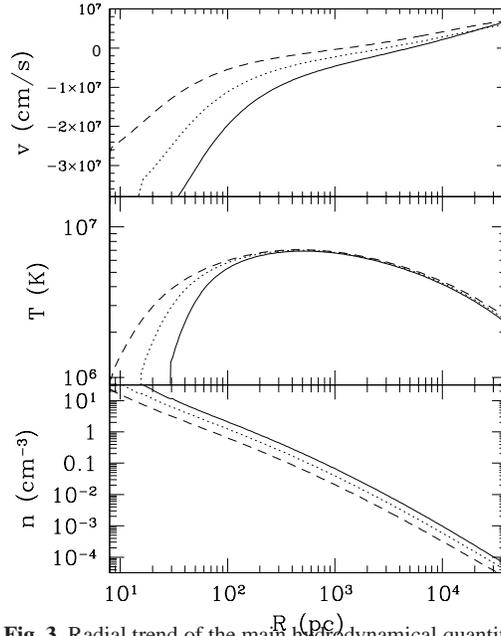}
\vskip -1truecm
\caption{Radial trend of the main hydrodynamical quantities of the
gas flow in the testcase ETG described in Sect.~\ref{test}: velocity (negative
inward), top;
temperature, middle; number density ($\rho /m_p$), bottom. The
galaxy age is 5 (solid), 8.5 (dotted) and 13 (dashed) Gyr.}
\label{f4}   
\end{figure}

\begin{figure}[t]
\vskip -3.5truecm
\includegraphics[scale=.45]{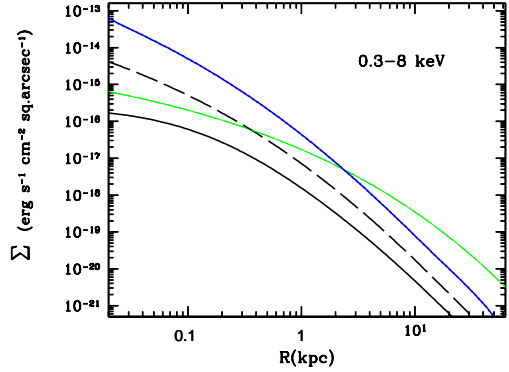}
%
%
\vskip -0.5truecm
\caption{The surface brightness profile for the model in
Fig.~\ref{f4} (blue line), and for two other models whose
hydrodynamical quantities are shown in the next Fig.~\ref{f7}, with
the same line type, all at an age of 13 Gyr. The green line follows the
optical profile of the testcase ETG, and is normalized to 
the emission of unresolved 
X-ray binaries, as 20\% of their total collective luminosity
(the unresolved fraction is $<25$\% for a
sample of local ETGs observed with $Chandra$, excluding 
the very hot gas rich ones; \cite{Kim10}).}
\label{f6}   
\end{figure}

The gas velocity, temperature and density profiles at representative
epochs are shown in Fig.~\ref{f4}; [...]
the 0.3--8 keV surface brightness profile at the
present epoch is shown in Fig.~\ref{f6}.  As typical of steep mass
models, a central inflowing region is present from the beginning; due
to the secular increase of the specific heating of the gas
(Eq.~\ref{eq:lsnspec}, Sect.~\ref{reasons}), 
with time  increasing the infall velocity 
decreases in modulus, the stagnation radius migrates inward 
(Fig.~\ref{f4}). By the present epoch a
quasi-stationary configuration establishes, by which the gas leaves
the galaxy at a rate almost equal to the rate $\dot M_*$ at which gas
is injected by stars, and the small difference goes into the central
sink. 

The final gas temperature profile [...] decreasing
outward is common among ETGs observed with $Chandra$ (e.g.,
\citep{Matsu,Fuk,DS2}); other profiles often occurring are flat or
outwardly increasing ones (\cite{DS2}, see also Statler, this
volume). To obtain the latter two shapes, other ingredients are 
required with respect to those included here, as AGN feedback
depositing heating outside the central galaxy core,  through a jet
or rising bubbles (e.g., \citep{Fin,For05,DS3,OB}), or as an external
medium (\citep{SW87,BT,BM98,DS3}).

(abridged)

\begin{figure}[t]
\includegraphics[scale=.5]{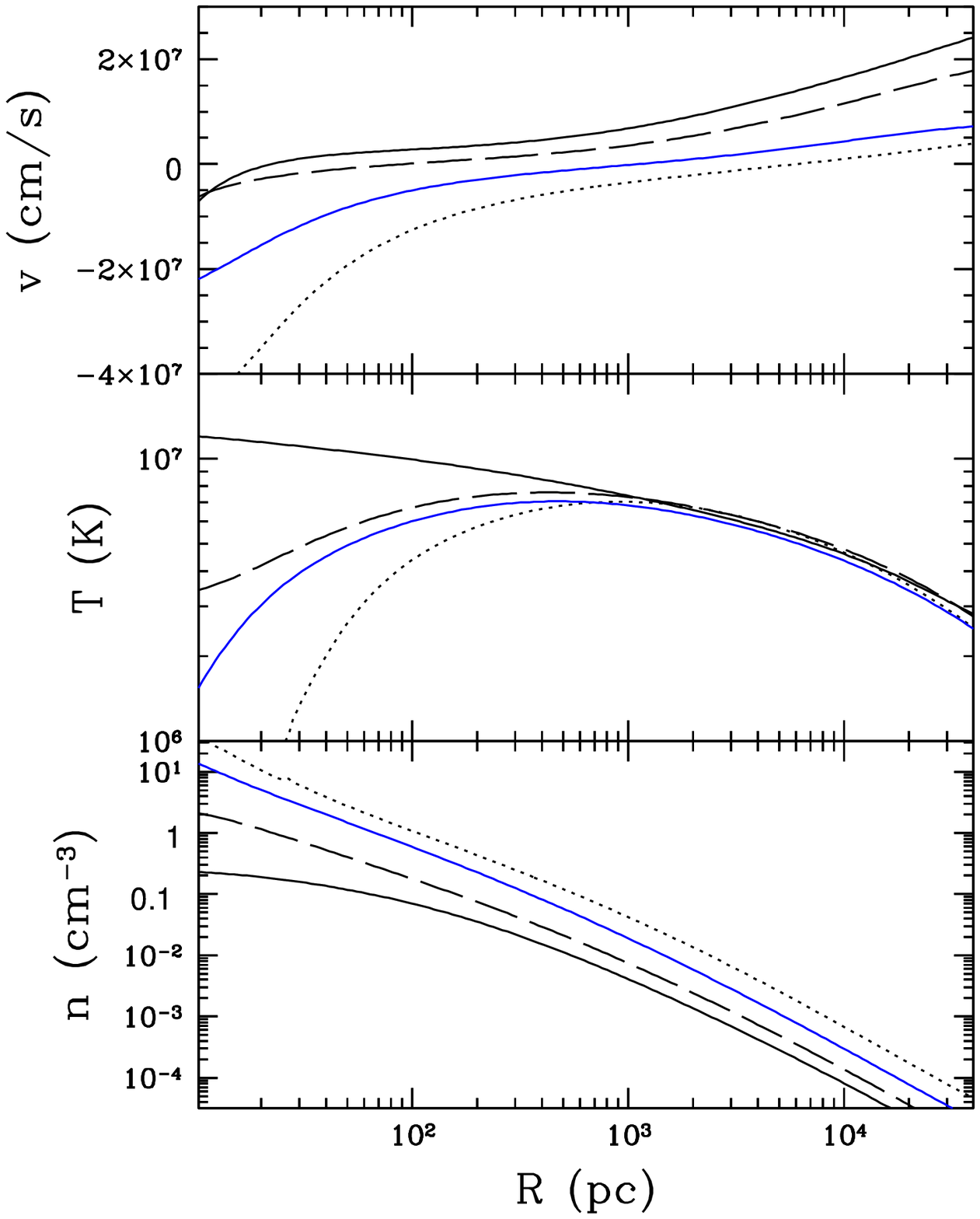}
\vskip-1truecm
\caption{Radial trend of the main hydrodynamical quantities at a
galaxy age of 13 Gyr, for the testcase ETG in Fig.~\ref{f4} (blue
lines), and for three galaxy models with the same $L_B$ and $\sigma_c$
except for the following differences: a SNIa's rate increased by 20\%
(dashed line); $n=6$, $M_*/L_B=6.1$ (solid
line); $n=4$, $M_*/L_B=8.5$ (dotted line).  The
corresponding gas luminosities are respectively
$L_{X,ISM}=1.1\times 10^{39}$, $4\times 10^{38}$,  
$3.0\times 10^{40}$ erg s$^{-1}$.  The brightness profiles of the
models shown with a dashed and solid line are plotted in Fig.~\ref{f6} (see
Sect.~\ref{test}).}  
\label{f7} 
\end{figure}

\subsubsection{Variations in the Testcase ETG and in $L_{X,ISM}$}\label{var}

See the full chapter (chapter 2 in {\it Hot Interstellar Matter in
Elliptical Galaxies},  Springer, 2012;
http://www.springer.com/astronomy/book/978-1-4614-0579-5)

\section{The Nuclear Scale}\label{mbh}

Accretion to the center is commonly present, though from a
small region, for the models in Fig.~\ref{f7}; therefore, this
Section explores what modifications to the flow are expected from the
addition of a central MBH. [...]  It is now clear that MBH
feedback is unavoidable on cosmological timescales; however, for
timescales much shorter than the cosmological one, and closer to the
present epoch, it is not fully understood yet how it works in
detail. It is then interesting to consider a few basics aspects of the flow
behavior for models with a central MBH but neglecting feedback, as:
how the flow is affected by the MBH gravity, the similarity of the
accretion flow with a Bondi flow, how much accretion energy is
expected, and how the picture outlined in the previous Sections
is modified.

\begin{figure}[t]
\hskip -1.truecm
\includegraphics[width=0.69\linewidth,height=0.9\linewidth]{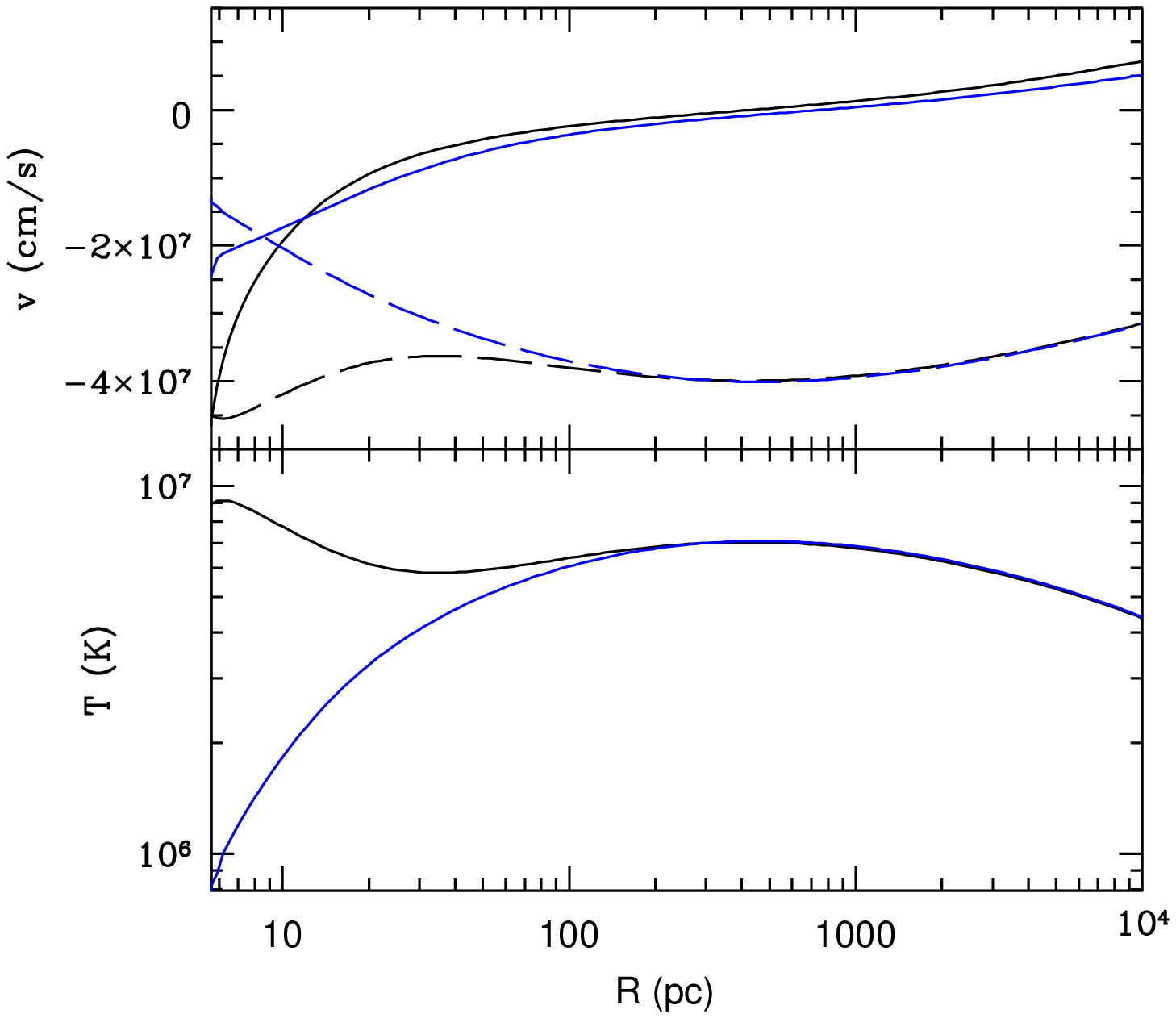}
\hskip -1.9truecm
\includegraphics[width=0.7\linewidth,height=0.9\linewidth]{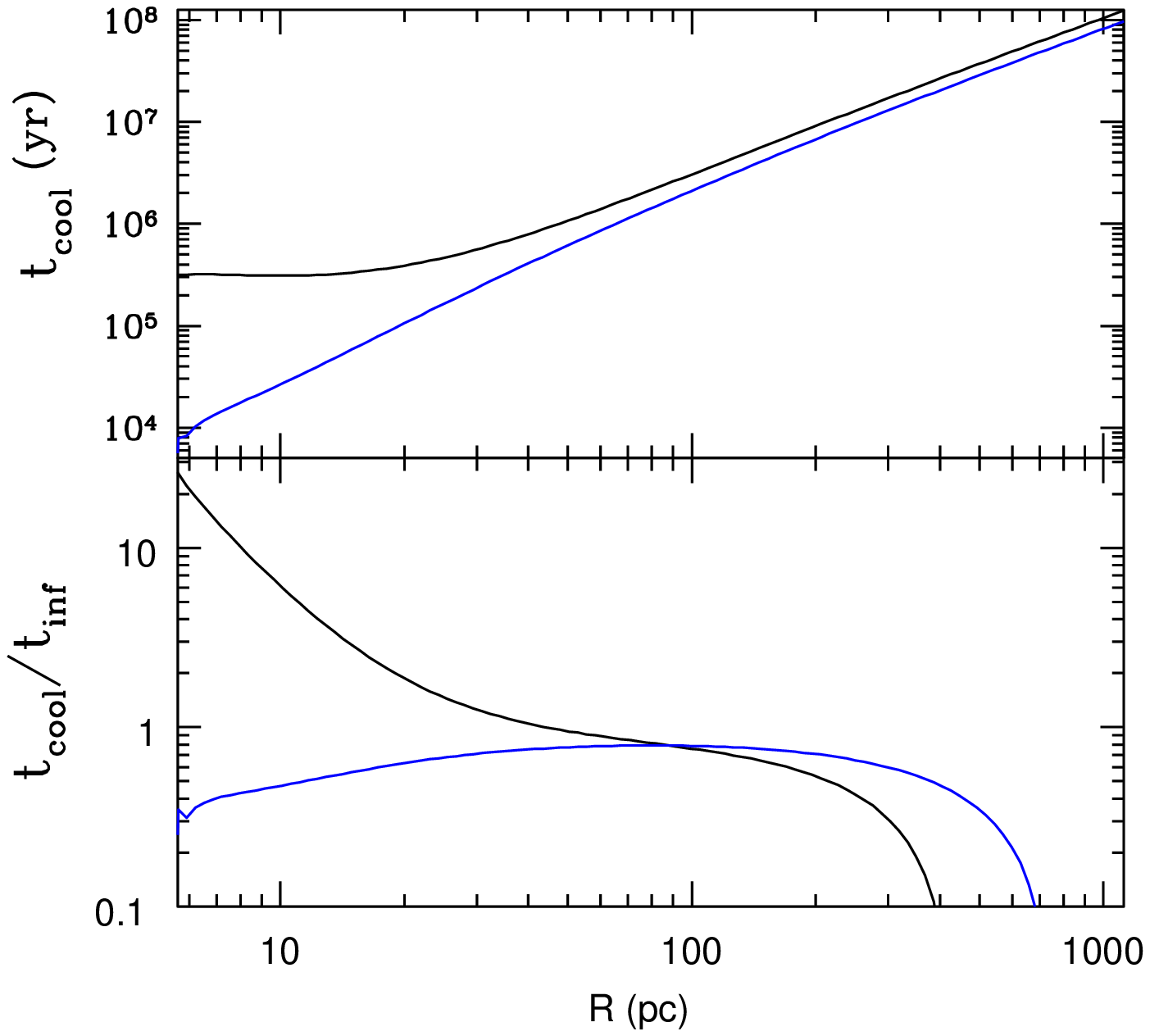}
\vskip-3.5truecm
\caption{Left: the velocity and temperature profiles for the testcase
ETG of Sect.~\ref{test} (in blue), and for the same galaxy
model with a central MBH of mass $3.6\times 10^8M_{\odot}$ added (in black), 
at an age of 13 Gyr; the dashed lines show $-v_{s}$
for $\gamma=5/3$, for the two models respectively.
Right: the cooling time and its ratio with the infall time, for the
models on the left. See Sect.~\ref{mbh} for more details.}
\label{f8}   
\end{figure}

\subsection{Gravitational Heating from the MBH  and Comparison with the
Bondi Accretion}

\begin{figure}[t]
\includegraphics[scale=.4]{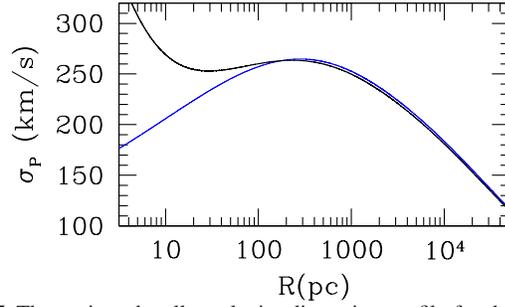}
\vskip-4.2truecm
\caption{The projected stellar velocity dispersion profile for the
galaxy model of the testcase ETG (Sect.~\ref{test}, dotted line), and for the
same model with a central MBH of mass $3\times 10^8M_{\odot}$
discussed in Sect.~\ref{mbh} (solid line).  Both models have an
aperture velocity dispersion of 260 km s$^{-1}$ within $R_e/8$.}
\label{f10}   
\end{figure}

The addition to the mass model of the testcase ETG (Sect.~\ref{test})
of a central gravitating mass $M_{BH}=3.6\times 10^8M_{\odot}$, as predicted
by the Magorrian relation, that remains constant with time, produces a
flow that is not influenced by the MBH in the outer region; within the
central $\sim 100$ pc, instead, the infall velocity and the
temperature become larger (Fig.~\ref{f8}), the density lower, and the
mass inflow rate to the center $\dot M_{in}$ smaller (from 0.12 to
0.06 $M_{\odot}$yr$^{-1}$ at a radius of 5 pc).  The most relevant
difference caused by the addition of the MBH is the higher gas
temperature at the center; this is due to the heating of the
gas produced by the sharp increase of the stellar velocity dispersion
within the sphere of influence of the MBH (of radius of the order of
$GM_{BH}/\sigma_c^2=20$ pc, where $\sigma_c$ is the velocity
dispersion without the MBH; Fig.~\ref{f10}), and by the compression of
the gas caused by the gravity of the MBH.  The MBH heating succeeds in
"filling" the central "hole" in temperature of the model without MBH,
and it may even be able to create a small central peak in the gas
temperature (Fig.~\ref{f8}).  The difference in central temperature
takes more time to establish for models with larger central gas
density (as that with $n=4$ in Fig.~\ref{f7}), and it may even not take place within the present epoch.  In
models where the flow, without the MBH, keeps hot down to small radii
(as that of the black solid lines in Fig.~\ref{f7}), the change in
temperature is far less dramatic (an increase by 25\% of the central
temperature), and $\dot M_{in}$
remains substantially unchanged at very small values ($\dot M_{in}\sim
10^{-4}-10^{-3}\,M_{\odot}$yr$^{-1}$).

Another important property of the flow is the value of its velocity
with respect to the sound velocity $v_{s}$, calculated for example for
the adiabatic case ($\gamma =5/3$; $v_s=\sqrt{\gamma kT/\mu m_p}$), and
also shown in Fig.~\ref{f8}.  Without a
MBH, in cases of "cold" accretion as for the testcase ETG, the flow
becomes supersonic close to the galactic center; with a MBH, accretion
is hot and keeps subsonic, while the flow tends to reach the sound
velocity at the innermost gridpoint (that, for this series of runs,
has been put to 1 pc). In fact, in absence of momentum feedback, it is
unavoidable for the flow to tend to the free fall velocity close to
the MBH, where the potential energy per particle becomes larger than
the thermal energy (and this is reproduced by the inner boundary
condition of a vanishing thermodynamical pressure gradient). 
The MBH heating also causes the cooling time $t_{cool}$ to become
much larger than the inflow time $t_{inf}$ within the central $\sim 100$ pc
(Fig.~\ref{f8}).  Both properties (the inflow
velocity that tends to $v_s$, and $t_{cool}>>t_{inf}$) characterize 
also the Bondi (1952) solution for spherically symmetric accretion on a
central point mass, from a nonrotating polytropic gas with given
density and temperature at infinity, in the adiabatic case ($\gamma
=5/3$).  This fact provides some support to a commonly used 
procedure to estimate the MBH mass accretion rate of ETGs (e.g., \cite{L01,So}),
that is the use of the analytic Bondi
(1952) formula, replacing infinity with a
fiducial accretion radius $r_{acc}=2GM_{BH}/v_s^2$ (\cite{Fra}), and
calculating $v_s$ as close as possible to the MBH. This is not a
trivial aspect since there are additional ingredients in the galactic flow
that are not included in the Bondi (1952) analysis, but are accounted
for by the simulations, as: 1) the presence of mass and energy sources, as
the stellar mass losses and the SNIa's heating; 2) the possibility of
cooling; 3) the fact that $r_{acc}$ is not a true infinity point,
since the gas experiences a pressure gradient there; 4) the
contribution of the galactic potential added to that of the MBH.
The simulations however have some limits too: for example, the
discrete nature of the stellar distribution becomes important where
the accretion time on the MBH ($t_{inf}\sim 10^4- 10^5$ yrs from 10
pc, in the simulations) is comparable to (or lower than) the time
required for the stellar mass losses to mix with the bulk flow
(\cite{Mat90,PB}), or to the time elapsing between one SNIa event and the
next (see also \cite{T10}).  Another limit is that some form of
accretion feedback is also expected, as briefly outlined in the next Section.

\subsection{The Importance of the Energy Output from Accretion}\label{imp}

The mass inflow rates of the models in Fig.~\ref{f7}, with a central
MBH added, range from $\dot M_{in}=4\times 10^{-4}$ (model with solid
line) to $4\times 10^{-1}$ $M_{\odot}$yr$^{-1}$ (model with dotted
line), at an inner gridpoint of 5 pc, at the present epoch.  If
totally accreted, the largest of these $\dot M_{in}$ releases an
accretion power $L_{acc}\sim 0.1 \dot M_{in} c^2 =2\times 10^{45}$ erg
s$^{-1}$ (\cite{Fra}), a large value that can have a significant
impact on the surrounding hot ISM, depending on the fraction that can
interact with the ISM and be transferred to it.  $L_{acc}$ is mostly
in radiative form at high mass accretion rates; more precisely, the
radiative efficiency of the accreting material can be written in a
general way as $\epsilon =0.2\times 100 \dot m/(1+100\dot m)$, where
$\dot m=\dot M_{in}/\dot M_{Edd}$ is the Eddington-scaled accretion
rate, and $\dot M_{Edd}=L_{Edd}/0.1 c^2=2.2\times
10^{-8}M_{BH}(M_{\odot})\,\,\, M_{\odot}$yr$^{-1}$ (\cite{CO10}). When
$\dot m \gsim 0.01$, then $\epsilon \sim 0.1-0.2$, and $L_{acc}$ is
mostly in radiative form; when $\dot m\ll 0.01$, then $\epsilon \sim
20 \dot m$, as for radiatively inefficient accretion flows
(\cite{Nar}), and the radiative output becomes negligible.  In the
low-$\epsilon$ regime, the output of accretion may be dominated by a
kinetic form (\cite{BB,Koe,All,Mer}).  For example, from the energy
input by accretion feedback to the hot coronae of a few nearby ETGs,
with $\dot M_{in}$ calculated using the Bondi rate
as described in Sect.~\ref{mbh} and $\dot m\ll 0.01$, it was found
that $\sim 1/5$ of $L_{acc}$ is converted into jet power
(\cite{All}).  The lowest $\dot m $ among the models in Fig.~\ref{f7}
is $\sim 10^{-4}$, and then accretion is highly radiatively
inefficient, but the largest is $\dot m =0.04$, [...].

For a representative model [obtained with the high resolution
simulations with radiative and mechanical feedback of Ciotti et
al. (2010), similar to the testcase ETG (Sect. 2.3.3)],
the duty cycle (the fraction of the time the AGN is
in the "on" state) is of the order of $10^{-2}$, for the past 5-7
Gyrs; outside the nuclear bursts, the flow behavior is similar to that
described in Sect.~\ref{dec} (\cite{CO10}), with a major difference:
the lower central gas density, due to the MBH heating
(\cite{PCO}). [...] The brightness profiles are then much less centrally peaked
than those in Sect. 2.3.3. [...]

In addition to this important (positive) effect on the surface
brightness profile of the hot gas, how does AGN feedback modify the
scenario outlined in the previous Sections? ETGs where
$L_{SN}>L_{grav}^-$, and then already outflow-dominated, will not be
affected by further sources of heating.
For the other ETGs, the answer depends on how
much energy from accretion is transferred to the hot ISM:
if this energy is $>> L_{SN}$ then the scenario above will be modified,
while if it is $<<L_{SN}$ it will be preserved.  In general, it can be noted
that the gas modeling based on realistic stellar and dark mass
profiles, stellar mass loss and supernova rates and their secular
evolution, without accretion feedback can already reproduce reasonably
well the fundamental gas properties (e.g., trend of $L_{X,ISM}$ with $L_B$, wide
variation in $L_{X,ISM}$, average gas temperature), therefore such
modeling must catch the bulk of the origin and evolution of the hot
gas in ETGs.  Moreover, even in the context of feedback modulated gas
flow evolution, the hot gas
content at the present epoch, seem still sensitive to the structural
galaxy parameters, in the same sense as described in
Sects.~\ref{enbud} and~\ref{dec} (Ciotti \& Ostriker, this volume; \cite{PCO}).
Finally, the modeling without feedback -- if any -- shows the need for gas accretion
from outside or confinement (Sect.~\ref{dec}); the nuclear energy
input should then mostly readjust the internal gas structure, without
causing major degassing at later epochs. The measure in which activity
affects the gas content is yet to be established observationally; so
far, exploiting $Chandra$ resolution, it has just been shown that the
nuclear X-ray luminosities of ETGs correlate only weakly with their gas
luminosity (\cite{P10}).

\section{Gas Flows and Galactic Shape, Rotation, Stellar Profile}\label{sensit}

In Sect.~\ref{dec} it was shown how the gas content of an ETG is
sensitive to changes regarding the stellar and dark mass components
that are in fact allowed for by observations (see, e.g., the scatter
around the fundamental scaling laws of ETGs), and by modeling (see,
e.g., how model ETGs lying on the Fundamental Plane can be built with
different $\cal{R}$ and $\beta $, \cite{RC93} and Sect.~\ref{var}).
This holds even at fixed $L_B$, so to account for a significant part
of the large observed $L_{X,ISM}$ variation. Below we consider 
the effects on the hot gas content produced by additional 
variations in the galactic structure that are observed and
have not been considered
above, such as the galactic shape, the amount of rotation in the
stellar motions, and the central stellar profile.

\begin{figure}[t]
\vskip -3truecm
\includegraphics[scale=.45]{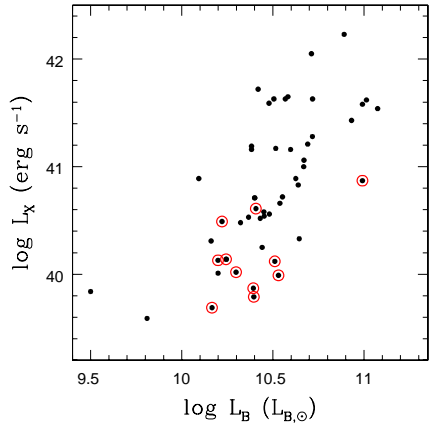}
\vskip -1truecm
\caption{The total X-ray emission (from $ROSAT$ PSPC observations, \cite{OS})
versus the galactic blue luminosity $L_B$ (from total B-magnitudes in the
Hyperleda catalog and distances in \cite{OS}) for a sample of ETGs
with measured optical flattening  and stellar rotation (\cite{P97}).  
Galaxies with high flattening ($b/a<0.6$)
are marked with a red circle (see Sect.~\ref{shape}
for more details).}
\label{lxlbba}   
\end{figure}

\subsection{Galactic Shape and Rotation}\label{shape}

Soon after the first large sample of ETGs with known X-ray emission
was built from $Einstein$ observations, it was found that the hot gas
retention capability is related to the intrinsic galactic shape: on
average, at any fixed $\lb$, rounder systems show larger total X-ray
emission $\lx$ and $\lx/\lb$ than flatter elliptical and S0 systems
\cite{Esk}.  The relationship defined by $\lx/\lb$ is stronger than
that defined by $\lx$. Moreover, galaxies with axial ratio $b/a$ close to
unity span the full range of $\lx$, while flat systems all have
$\lx\lsim 10^{41}$ erg s$^{-1}$ (see, e.g., Fig.~\ref{lxlbba}).  A
similar result holds for the "diskiness" or "boxiness" property of
ETGs, that measures the deviation of the isophotal shape from a pure
elliptical one (\cite{Ben,KB}). This property is described by the $a_4/a$
parameter, in a way that disky ($a_4>0$) ETGs show isophotes distorted
in the sense of a disk, and boxy ($a_4<0$) ETGs have isophotal
distortions in the sense of a box.  Disky systems are also generally
flattened by rotation, while boxy ones have various degrees of
velocity anisotropy (see also \cite{Pas}).  Boxy ETGs cover the whole
observed range of $\lx$, while disky ETGs are less X-ray luminous on
average (\cite{Ben,Esk}); this result is not produced only by disky galaxies
having a lower average galactic luminosity, with respect to boxy ETGs,
since it holds even in the range of $L_B$ where the two types overlap
(\cite{P99b}).  The relationship between $\lx$ and $a_4/a$ was
reconsidered, confirming the above trends, for the $ROSAT$ PSPC sample
(\cite{EO}).

\begin{figure}[t]
\vskip -3truecm 
\includegraphics[scale=.45]{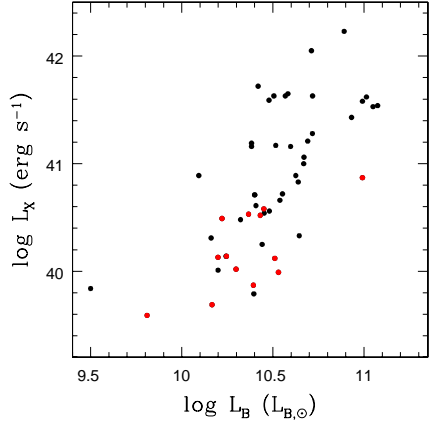}
\vskip -1truecm
\caption{The total X-ray emission versus the galactic luminosity $L_B$
(both determined as in Fig.~\ref{lxlbba}); galaxies with $\vr/\sgc>0.5$
are marked with a red symbol (see Sect.~\ref{shape} for more details).}
\label{lxlbrot}   
\end{figure}

There seems then to be a dependence of the hot gas content on the
galactic shape, measured by either $\epsilon =1-b/a$ or $a_4/a$.
Since flatter and disky systems also possess, on average, higher
rotation levels (\cite{bt87}), the influence on the
hot gas of both the shape of the potential well and of the stellar
rotation was called into question.  The gas in ETGs that are in part
rotationally supported may have a lower ``effective'' binding energy
per unit luminosity compared to the gas in non-rotating ones (a lower
effective $L_{grav}^-/\lb$ ratio, in the notation of
Sect.~\ref{enbud}; see also Sect.~\ref{sharot} below), and then
rotating ETGs may be more prone to host outflowing regions.  For this
reason, the effects on $\lx$ of the ellipticity $\epsilon$ of the
stellar distribution and of stellar rotation were studied for a sample
of 52 ETGs with known $\lx$, maximum rotational velocity of the stars
$\vr$, and central stellar velocity dispersion $\sgc$ (\cite{P97}).
[...]p The gas
content can be high only for values of $\vr/\sgc\lsim 0.4$, while
modest or low gas contents, as log[$\lx$(erg
s$^{-1})/\lb(L_{B,\odot})]\lsim 30.2$, are independent of the degree
of rotational support.  Recently, for the ETGs of the $SAURON$ sample,
the relationship between soft X-ray emission and rotational properties
was investigated again (\cite{Sar}), confirming that slowly rotating
galaxies can exhibit much larger luminosities than fast-rotating ones.
As for the axial ratio and the isophotal parameter $a_4$, the trend of
$\lx/\lb$ with $\vr/\sgc$ is not produced by the correlation between
$\lx$ and $\lb$: ETGs with high $\vr/\sgc$ cover substantially the
same large range in $\lb$ as the whole sample (Fig.~\ref{lxlbrot} ).

In conclusion, rotation seems to have an effect similar to that of
shape, and $L_X$ and $L_X/L_B$ show a similar trend with respect to
axial ratio, diskiness, and rotation: their variation is large for
round, boxy and slowly-rotating systems, while it keeps below a
threshold for flatter, disky and high-rotation systems.  From
observations it remains then undecided which one between axial ratio,
diskiness, and rotation is responsible for the trend; more insights is
given by the theoretical and numerical analysis discussed below.

\subsubsection{A Theoretical and Numerical 
Investigation }\label{sharot}

The impact of stellar rotation and galactic shape on the hot gas content was
also addressed with theoretical and numerical studies
(\cite{CP96,BM96,DC}).  In principle, the lower gas content of flatter
systems could be due to the mass distribution itself, or to 
a higher rotational level that decreases the
effective potential.  An
analytic investigation showed that flatter systems are
less able to retain hot gaseous halos than rounder ones of the same
$\lb$, due to the effect of the shape more than that of a larger
rotational level (\cite{CP96}). The investigation reconsidered the global estimate of
the energy budget of the gas introduced in Sect.~\ref{enbud},
generalizing it for flows in flat and rotating galaxy models.  The
classical scalar virial theorem for a stellar distribution interacting
with a dark matter potential can be written as $2T+\Pi=|W|$, where
$\Pi$ and $T$ are the kinetic energies associated respectively with
the stellar random\footnote{In the notation used here and in
Sect.~\ref{enbud}, $\Pi$ is twice the energy due to random motions.}
and ordered motions, and $|W|$ is the potential energy of the stellar
component plus the virial interaction energy of the stars with the
dark halo.  For a fixed total mass and mass distribution (i.e., a
fixed $|W|$), the amount of rotational streaming energy $T$ can
formally vary from zero to a maximum that depends on the galaxy
structure (\cite{CP96}); in the notation of
Sect.~\ref{enbud}, the power $L_{rot}$ related to rotational streaming
is $L_{rot}=\dot M_*\, T/M_*$, while that related
to random motions is $L_{\sigma}=0.5\dot M_*\, \Pi /M_*$.
How does $L_{rot}$ enter the energy budget of the gas, for
example in Eq.~\ref{eq:chi}, for a fixed $|W|=2\, M_* \,
(L_{rot}+L_{\sigma})\, /\, \dot M_*$ ? In an extreme case, the whole effect
of the ordered motion is to produce a change in the effective potential
experienced by the gas, if for example the galactic corona is rotating with the
same rotation velocity as the stars; in this case $L_{rot}$ is to be
subtracted from $L_{grav}^-$. In the opposite extreme case, all the
kinetic energy of the gas, from random and from ordered motions, is
eventually thermalized; then $L_{rot}$ is to be added to $L_{SN}$ and
$L_{\sigma}$, in the denominator of Eq.~\ref{eq:chi}. The real
behavior, lying between the two extreme cases, can be parameterized
re-writing Eq.~\ref{eq:chi} as
\begin{equation} 
{L_{grav}^- -\alpha L_{rot}\over L_{SN} + L_{\sigma}
+(1-\alpha )L_{rot}} 
\label{eq:lrot} 
\end{equation} 
with $0\leq\alpha\leq 1$. If $\alpha =0$, the thermalization of
$L_{rot}$ is complete, and since the kinetic energy of stellar motions
($L_{\sigma}+L_{rot}$) will be lower\footnote{For a totally
velocity dispersion supported galaxy, $L_{\sigma}$ accounts for the
whole energy input to the gas from the stellar motions, that is significantly
lower than $L_{SN}$ (Sect.~\ref{enbud}, below Eq.~\ref{eq:lsig}).}
than $L_{SN}$, then Eq.~\ref{eq:lrot} coincides with Eq.~\ref{eq:chi}.
If instead $\alpha =1$, there is no thermalization of $L_{rot}$, the
decrease of $L_{grav}^-$ is maximum, and the effect of rotation is
maximum.  However, it is found that the role of rotation remains
minor, because it can change Eq.~\ref{eq:lrot} by only a few per cent:
the variation of Eq.~\ref{eq:lrot},
between the null and the maximum $L_{rot}$ allowed by realistic galaxy models, 
is small, even for
$\alpha =1$ ($<10$\%; \cite{CP96}). Instead,  variations of
more significant amount that can make the gas significantly less bound (a
decrease in $L_{grav}^-$ of $\sim 20$\%) can be produced by a change
in the galaxy structure, as a reasonable flattening of a round system
at fixed $\lb$.  Therefore, S0s and non-spherical ellipticals are less
able to retain hot gaseous haloes than are rounder systems of the same
$\lb$, and more likely to host outflowing regions.

The results of the purely energetical approach above were tested with
numerical studies of gas flows.  Two-dimensional simulations for
oblate ETGs, with different amounts of ordered and disordered kinetic
energies, were carried out for gas in the inflow state (\cite{BM96}).
In this investigation $\lx$ is reduced in rotating models, because the
gas cools on a disk before entering the galactic core region, and then
$L_{grav}^+$ (Eq.~\ref{eq:lp}) is reduced; since rotation increases on
average with flatness, rotation would be the underlying cause of the
X-ray underluminosity of flat objects.  However, the massive,
rotationally supported, and extended cold disk that forms in the
equatorial plane, due to mass and angular momentum conservation, and
comparable in size to the effective radius, is not observed; also, the
resulting X-ray images should be considerably flattened towards the
equatorial plane out to an optical effective radius or beyond, a
phenomenon that is small or absent (\cite{Han}).  Other authors
(\cite{DC}) performed two-dimensional numerical simulations of gas
flows for flat systems, but allowing for the gas to be outflowing.
The flows then developed a partial wind in flat ETGs that, if
spherical,  would be in inflow.  In this way, the
models accumulate negligible amounts of cold gas on a central disk.
Rotation could also decrease the X-ray emission (of a factor of two or
less), because it favoured the wind. In this scenario, then,
flat models, rotating or not, can be significantly less X-ray luminous
than spherical ones of the same $\lb$, because they are in partial
wind when the spherical ones are in inflow; rotation has an additional
but less important effect.

\subsection{The Central Stellar Profile}\label{cusp}

[...] Interestingly, the radio luminosity $L_{R}$ shows the same
behavior as the total soft X-ray emission with respect to the inner
stellar light profile: cusp ETGs are confined below a threshold in
$L_{R}$, while core ones span a large range of $L_{R}$
(\cite{Ben,CB,P10}).  Core systems can then reach the highest $L_R$
and possess a conspicuous radio activity cycle, while in cusp galaxies
the radio emission keeps smaller, likely because of a rapid jet
failure due to the lack of a dense confining medium, or a smaller duty
cycle (\cite{P10} and references therein).

\begin{acknowledgement}
I acknowledge support from the Italian Ministery of Education,
University and Research (MIUR) through the Funding Program PRIN 2008.

\end{acknowledgement}

\end{document}